\documentclass{article}

\usepackage[utf8]{inputenc}
\usepackage[T1]{fontenc}
\usepackage{lmodern}
\usepackage{setspace}
\usepackage{geometry}
\usepackage{titlesec}
\usepackage{xcolor}
\usepackage{graphicx}
\usepackage{lipsum}

\newtheorem{definition}{Definition}
\usepackage{amsmath}
\usepackage{float}
\usepackage{hyperref}

\definecolor{titlecolor}{RGB}{0,0,0} % Black
\definecolor{sectioncolor}{RGB}{44,62,80} % Dark blue
\definecolor{subsectioncolor}{RGB}{52,152,219} % Blue
\definecolor{linkcolor}{RGB}{52,152,219} % Blue

\titleformat{\section}
  {\color{sectioncolor}\normalfont\Large\bfseries}
  {\color{sectioncolor}\thesection}{1em}{}

\titleformat{\subsection}
  {\color{subsectioncolor}\normalfont\large\bfseries}
  {\color{subsectioncolor}\thesubsection}{1em}{}

\title{IdentityChain}
\author{Mahdi Darabi \\ \href{mailto:mahdi.darabi.official@gmail.com }{mahdi.darabi.official@gmail.com } 
\and Amirreza Fathi \\ \href{amirreza.f.amirreza@gmail.com}{amirreza.f.amirreza@gmail.com}}

\begin{document}
\maketitle

% Title
% \begin{center}
    % \Huge\bfseries\color{titlecolor}IdentityChain
% \end{center}

% Author and Date
% \begin{center}
%   \large\textit{Mahdi Darabi, Amirreza Fathi}
    
%   May, 2024
%\end{center}

% Abstract
\section*{Abstract}
The first generation of cryptocurrencies introduced revolutionary concepts, yet faced challenges in privacy and regulatory compliance. While subsequent cryptocurrencies aimed to address privacy concerns (like Zcash and Monero), they often conflicted with regulatory frameworks, hindering broader adoption. In response, inspired by recent researches about privacy and accountability \cite{damgaard2021balancing} and incentive techniques \cite{han2022can} in Blockchain, we propose IdentityChain as a novel framework that integrates privacy and accountability principles, leading to a robust system equipped with adaptable rules. 

IdentityChain is a KYC (Know Your Customer) service on top of a public Blockchain (e.g., Ethereum, Ton, Polygon). The goal is to maintain privacy while ensuring compliance with existing regulations. Privacy is one of the key characteristics of IdentityChain, it's crucial for preventing conflicts of interests further discussed how. Accountability is also one of the main characteristics of IdentityChain and prevents from misbehave of users. Privacy and accountability together wouldn't be possible unless advancements in cryptography. 
% The whole IdentityChain theoretically is based on some theories in the field of "Balancing Privacy and Accountability in Blockchain".
This paper discusses a system design with on-chain and off-chain components to implement a service with respect to privacy and accountability.  

% Table of Contents
% \newpage
\tableofcontents

% Introduction
\newpage

\section{Introduction}

Blockchain is a decentralized infrastructure, as a public ledger it's maintained by a network and doesn't rely on any trusted third party. Blockchain as a revolutionary technology has enabled many possibilities such as border-less transfers, smart contracts and many other applications specially in the field of finance (mostly known as DeFi). Although blockchain has enabled a completely new area of finance, but its complete capability hasn't been utilized yet. There are some barriers prevent from utilizing the full capabilities of blockchain, and one of the most important ones is regulatory. In many countries there's a regulatory concern about using blockchain and keeping users accountable. If a project can guarantee privacy of users and keep them accountable at the same time, then it can make a whole new area with minimum regulatory concerns. IdentityChain aims to build a novel framework guaranteeing both privacy and accountability, in next sections we'll see how these two characteristics play a key role to secure the system and eliminate conflict of interest between different parties of the system. 

The current landscape of identity management predominantly relies on centralized systems, which pose significant privacy risks and potential conflicts of interest. IdentityChain offers a solution by empowering users with control over their identity data while enhancing both privacy and accountability. However, achieving this balance in decentralized systems presents a challenge, particularly when it comes to aligning privacy with regulatory requirements and ensuring user accountability. IdentityChain addresses these challenges by utilizing advanced cryptographic techniques and distributed ledger technology. These innovations enable users to manage their identities without relying on a single, centralized authority. By decentralizing control, IdentityChain minimizes the risks associated with data breaches and unauthorized access. Moreover, IdentityChain ensures compliance with regulatory frameworks through robust mechanisms that allow for the verification and auditability of user actions without compromising individual privacy. This approach not only protects users' sensitive information but also deters malicious activities by maintaining a transparent and accountable environment.

IdentityChain will provide a whole new class of applications on blockchain, including but not limitted to: \begin{itemize}
\item 
Single-sign-on is one of the services can be implemented on top of identityChain. The growth of financial technologies is recent years and maybe in next years, has shown that KYC service is very crucial. Nowadays many fintech companies do KYC by their own, and KYC is a barrier to entry for many fintech startups. Also outsourcing KYC services to a third party will expose to many risks such as user base theft and conflicts of interest. Imagine a service that with just one KYC you can use a variety of applications and with full privacy. 
\item 
Exclusive services are among the possible services of IdentityChain. Imagine some on-chain launch pads and ICOs or NFTs that only are accessible by some verified users, but the Identity of users are totally anonymous for everyone in the system.
\item 
Real world credit score is another interesting service would be possible by IdentityChain. A user can lock their assets in their bank and based on that use different services on blockchain.
\item 
IdentityChain will bring many use cases into reality. Privacy and accountability together can overcome many adoption barriers of a blockchain service, since total privacy will eliminate most of conflict of interest problems, and accountability will be aligned with the regulatory concerns. 
\end{itemize}

Our system is also fully compliant with international KYC standards for the fintech sector, such as those set by the Financial Action Task Force (FATF) and the European Union’s Anti-Money Laundering Directives (AMLD). By adhering to these rigorous standards, our solution ensures that it meets the regulatory requirements necessary for combating money laundering and terrorist financing while still prioritizing user privacy. The rapid growth of the fintech industry, which increasingly relies on secure and efficient KYC processes, underscores the importance of our work. As fintech continues to expand globally, the demand for innovative KYC solutions that offer both security and compliance is more critical than ever. Our system not only addresses these needs but also sets a new benchmark for trust and accountability in the fintech landscape.

The following sections are organized as follows: In Section \ref{Related_work}, we discuss related works. In Section \ref{System_design}, we explain the system model and demonstrate the overall process of the system components. In Section \ref{Privacy_and_accountability}, we describe the theoretical models that ensure privacy and accountability. In Section \ref{System_Setup}, we explain the system setup, including how data is written and modified in smart contracts. In Section \ref{Implementation}, we outline the vision and implementation requirements, and in Section \ref{Conclusion}, we provide the conclusion and suggest future directions.

\section{Related work}
\label{Related_work}
The traditional KYC process for DeFi companies and banks is often time-consuming, inefficient, costly, and does not provide adequate data security and privacy. It also lacks a competitive environment for service improvement and faces challenges with outsourcing. However, the rapid growth of new companies and the increasing need for efficient customer acquisition and service delivery have created a demand for intelligent and automated KYC solutions \cite{schar2021decentralized}. As a result, various research efforts have been undertaken to address these challenges. For example, the authors in \cite{kumar2020blockchain} propose a blockchain-based cryptographic system in which user data is securely stored and can only be accessed in a distributed environment with the user’s authorization. In \cite{hajiabbasi2023cyber}, a design is introduced where biometric data is stored on the blockchain using asymmetric cryptography for authentication purposes in banks, including mechanisms for running neural models on this data. In \cite{druagan2020kychain}, the authors propose a hybrid system that stores data both on and off the blockchain, enabling users to manage and monitor their data while ensuring privacy and compensating various stakeholders involved. 

Blockchain plays a significant role in automating the KYC process, as it can advance procedures with the help of smart contracts without the need for a central authority. Additionally, cryptographic tools are available to ensure system security and maintain user privacy in this environment. Another important aspect of blockchain is its ability to facilitate payments, penalties, and commitments, which ensure the enforceability of algorithms. Moreover, the blockchain environment is transparent and scalable, making it suitable for various applications. Wallets and monitoring tools are also available in this space, making development and implementation more feasible. For an overview of some research in this direction, refer to studies \cite{hannan2023systematic} and \cite{malhotra2022blockchain}.

One of the challenges in blockchain systems is ensuring user accountability while maintaining anonymity. Legal authorities are concerned about preventing criminal activities and have made various efforts to monitor blockchain-based systems \cite{alves2024enhancing, maione2024blockchain, akanfe2024blockchain}. In this context, numerous academic efforts have been undertaken to address various law enforcement requirements using cryptographic tools. For instance, article \cite{li2019traceable} explores methods to make the Monero blockchain \cite{citekey1}, where users remain anonymous, traceable, thereby enabling legal authorities to investigate suspicious transactions. In article \cite{gao2024aassi}, the authors have designed and implemented an identity management system where users receive disposable identities. In this blockchain-based system, it is possible for legal authorities to selectively disclose user identities. Additionally, the system allows for various methods of tracing user activities. In article \cite{cheng2024s}, a system based on consortium blockchains is presented, where distributed identity authentication is performed, and user identities can be disclosed by regulators. This work also emphasizes the efficiency of algorithms in terms of computational costs. In article \cite{li2024idea}, methods for automatic digital identity verification are presented where identities based on JSON are issued. Users can authenticate themselves with these certificates at external service providers without needing to refer back to the issuer. An interesting point in this article is the ability to edit these identities, designed using zero-knowledge proofs for verifiability.

In this article, we adopted article \cite{damgaard2021balancing} as an off-the-shelf method. Similar to the aforementioned articles, our work incorporates features such as identity verification, anonymous user functionality, and selective disclosure of identity. Additionally, we considered performance requirements within the Ethereum blockchain \cite{kushwaha2022ethereum, soud2024fly, soud2024soley} and introduced IdentityChain by designing incentive and penalty mechanisms \cite{turan2024semi, karakostas2024blockchain, mssassi2024game}. In IdentityChain, users undergo identity verification across various geographical regions and intelligently and securely access desired services while preserving privacy. IdentityChain can align with diverse local and global regulations to combat money laundering and adhere to KYC requirements \cite{mekpor2018determinants, muradyan2022efficiency, network2022re}.

% Solution
\section{System Design}
\label{System_design}

\begin{figure}[H]    
    \centering
    \includegraphics[width=1\textwidth]{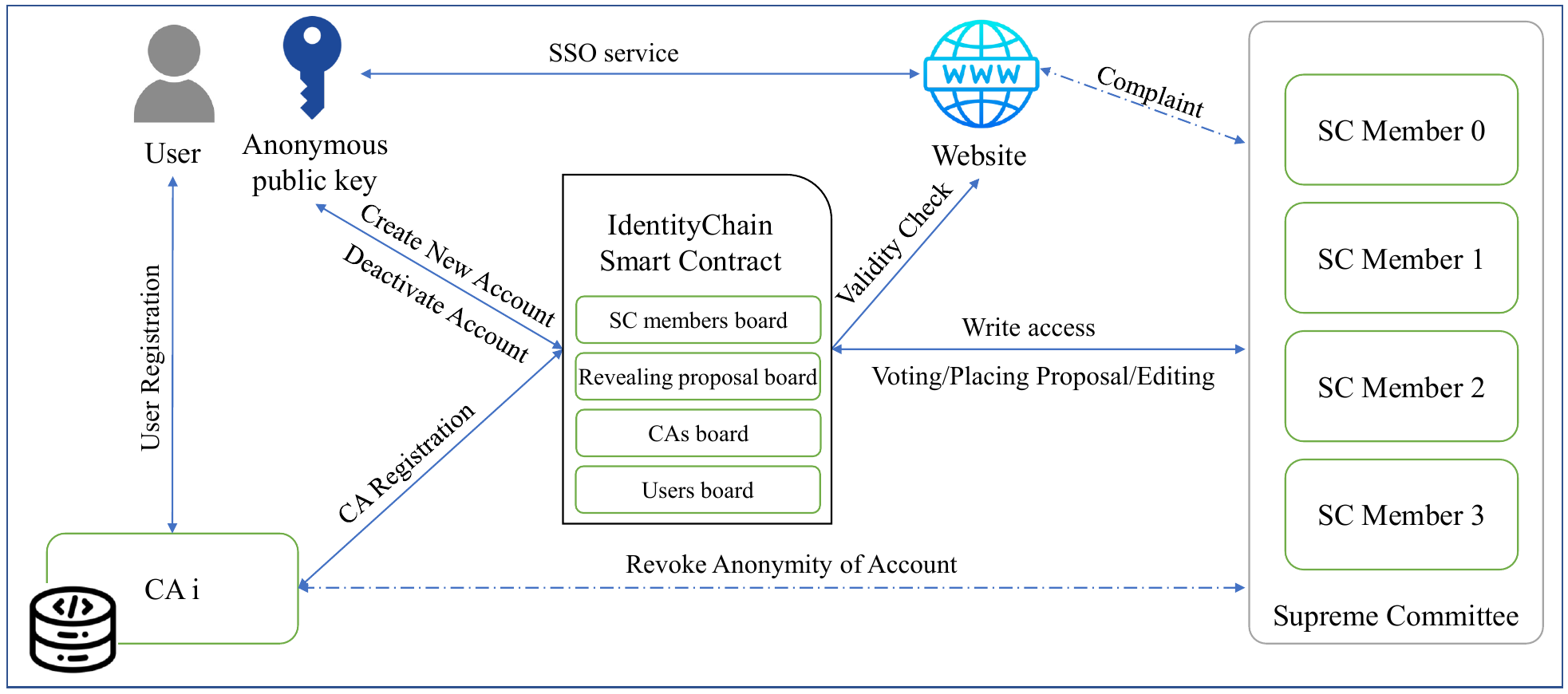}   
    \caption{Components of the IdentityChain system and their interactions.} 
    \label{System_model}
\end{figure}

We have designed the IdentityChain system, where websites can delegate their KYC processes. In this system, users undergo identity verification and receive a certificate that allows them to create a verified account and then request access to various websites. Some advantages of this system include:
\begin{itemize}
\item 
All processes are recorded on the blockchain transparently and immutably.
\item 
User confidentiality is maintained, and no one can identify a website’s customers by accessing the blockchain.
\item 
A committee comprising regulatory bodies can disclose user identities when necessary.
\item 
All components operate without a central authority, with mechanisms for rewards, penalties, and system exit processes in place.
\item
The system's structure is flexible concerning various regulations and allows for the implementation of new rules.
\end{itemize}
An overview of this system, its main components, and their interactions are shown in Figure \ref{System_model}.

In this section, we provide an overview of the tasks and responsibilities of the various entities within the system and explain how they collaborate. While this overview may seem vague, more detailed explanations are provided in Section \ref{Privacy_and_accountability} for clarification.

Supreme Committee (\texttt{SC}):
The Supreme Committee (\texttt{SC}) is an integral part of the system, with one of its primary responsibilities being to hold users accountable. This committee is responsible for various duties, including but not limited to: (1) voting on the admission of new SC members, (2) deciding on the acceptance or rejection of new Certificate Authorities (\texttt{CA}s), and (3) voting on whether to approve or reject the disclosure of a \texttt{User}'s identity.

It should be mentioned that a Revealing Proposal (\texttt{RP}) is a suggestion made by an \texttt{SC} member to all other \texttt{SC} members to disclose a user's identity. This identity will only be accessible to \texttt{SC} members. \texttt{SC} members are stakeholders of IdentityChain, and they are incentivized to perform their tasks diligently to increase the value of their holdings. If any \texttt{SC} members misbehave, the others in the committee can expel them and burn their stake. 

\begin{figure}[H]    
    \centering
    \includegraphics[width=1\textwidth]{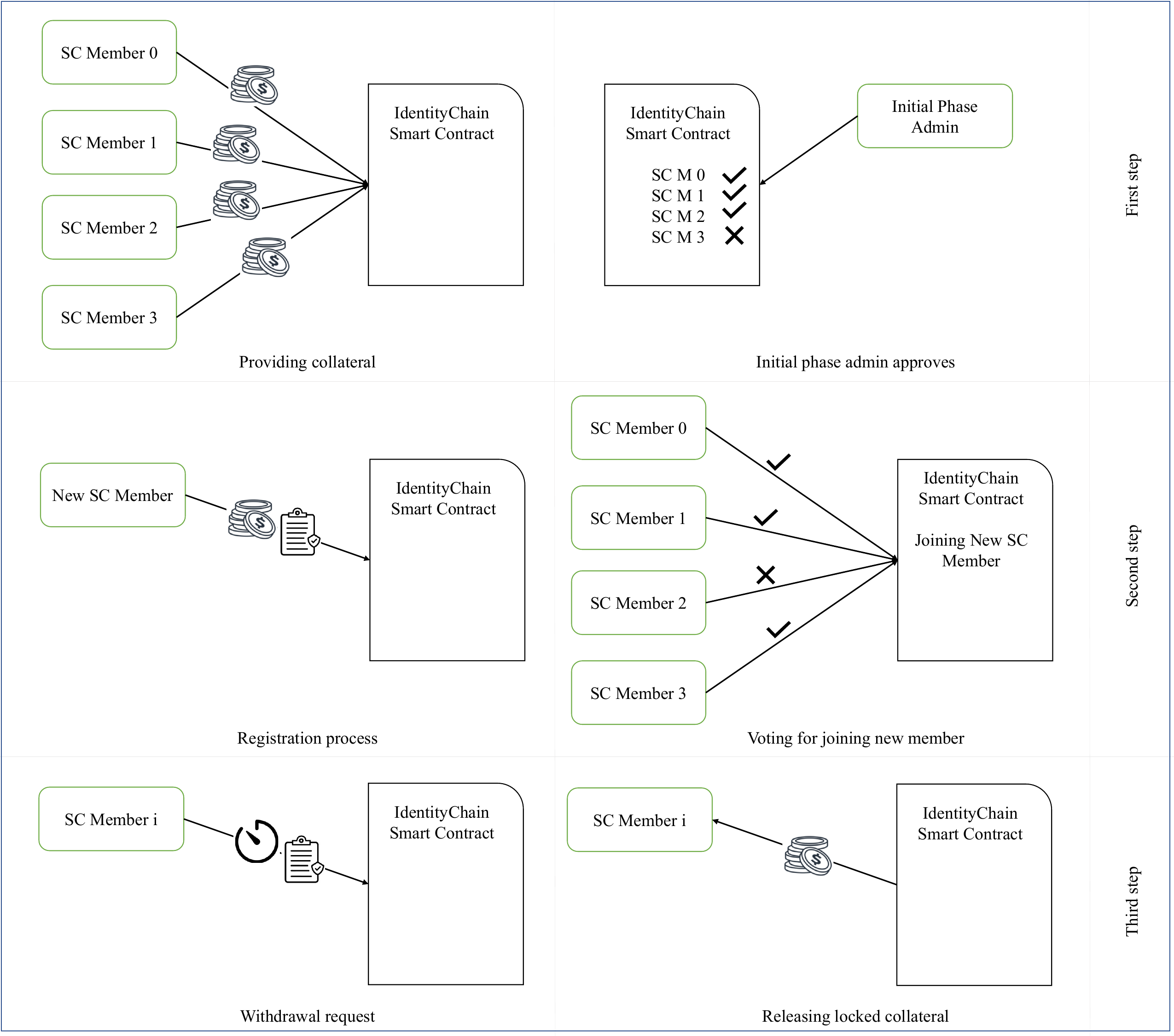} 
    \caption{This figure shows the three steps of \texttt{SC} member operations. Step one pertains to becoming a member during the initial phase. Step two pertains to a new individual becoming a member after the initial phase. Step three pertains to a member exiting the \texttt{SC}. \texttt{SC} members may vote to expel a member, but since this is not part of the normal process, it is not shown in this figure.}
    \label{SC_steps}
\end{figure}

An \texttt{SC} member can join the system through several steps:
\begin{itemize}
\item First step: At the beginning of the system a few semi-trusted parties provide some amount of money in IdentityChain smart contracts as collateral and become Supreme Committee member
\item Second step: Subsequently, anyone can join this committee through a defined process. This process entails the new member locking a certain amount of money, followed by a vote among the current Supreme Committee members regarding their admission. Depending on the voting outcome, the new member either joins the Supreme Committee or retrieves their locked funds, albeit with deductions for fees. 
\item Third step: If an \texttt{SC} member intends to exit the system, they must notify the other \texttt{SC} members at least six months in advance. Additionally, at the time of departure, there should be no ongoing protocols or pending votes requiring their participation. In this scenario, they can exit the system and free their stake. If an \texttt{SC} member exits the system without this coordination and fails to fulfill their responsibilities, the remaining \texttt{SC} members can vote to burn all or part of their stake.
\end{itemize}
The schematic of the three aforementioned steps is shown in Figure \ref{SC_steps}.

Certificate Authority (\texttt{CA}): 
The Certificate Authority (\texttt{CA}) plays a vital role in facilitating the KYC process within the system. In IdentityChain, users can only gain access by obtaining a certificate from a \texttt{CA}. But how does a \texttt{CA} become part of the system itself?

A \texttt{CA} can join the system through several steps:
\begin{itemize}
\item First step: The new \texttt{CA} must lock a certain amount of money as collateral and agrees to terms and conditions. 
\item Second step: In the subsequent step, a voting process is conducted, with Supreme Committee (\texttt{SC}) members voting on the \texttt{CA}'s inclusion. Depending on the outcome of the vote, the \texttt{CA} will either be admitted into the system or receive back its collateral minus fees. 
\item Third step: If a \texttt{CA} wishes to exit the system, it must notify the \texttt{SC} at least six months in advance of its intended departure date. Additionally, it must transfer its database to a new \texttt{CA}. Only then can it release its collateral after exiting the system. If a \texttt{CA} fails to properly respond to the \texttt{SC}'s requests, does not collect the necessary data from users according to KYC standards, or exits the system without coordination, its collateral will be burned.
\end{itemize}

\begin{figure}[H]    
    \centering
    \includegraphics[width=1\textwidth]{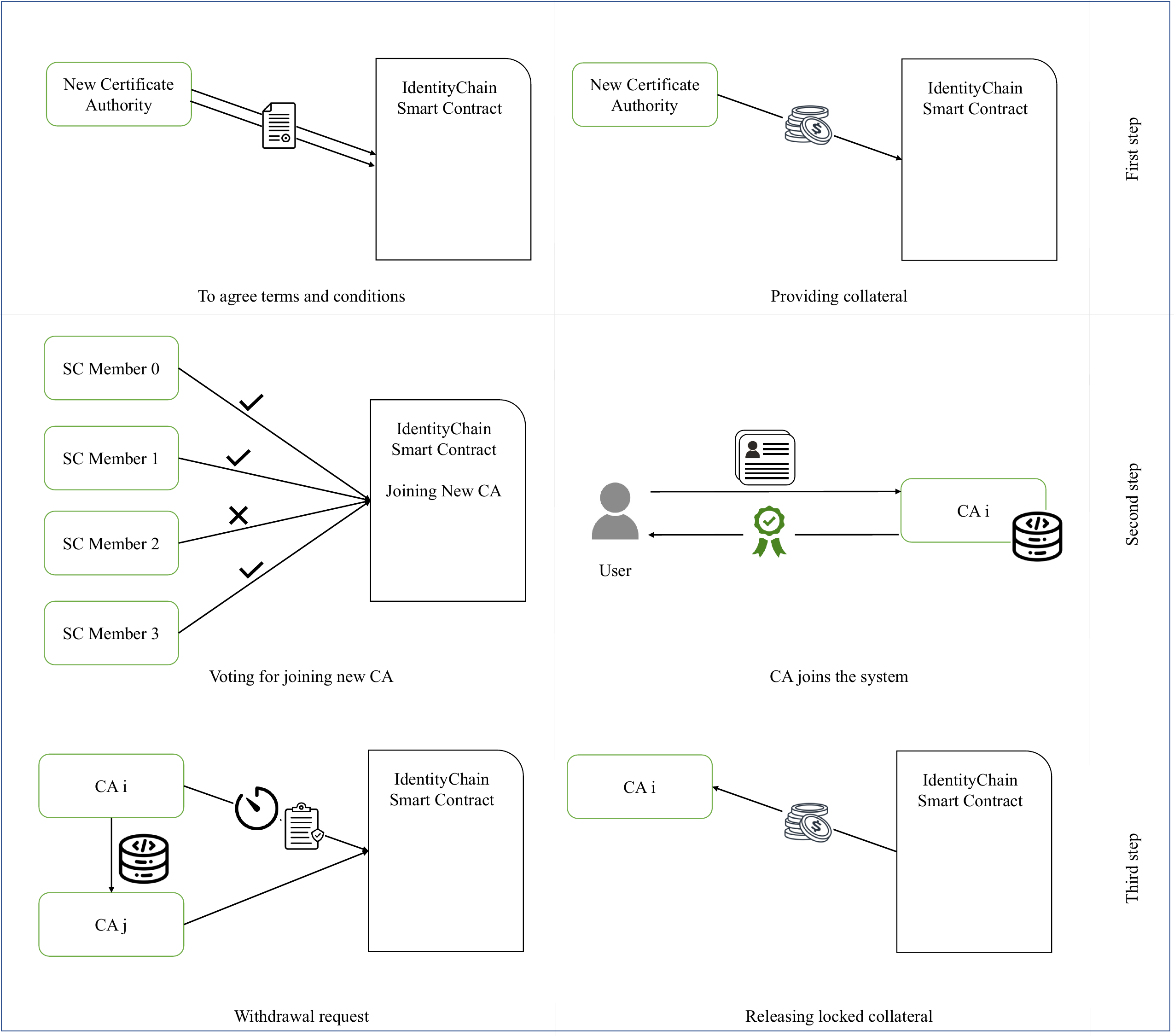} 
    \caption{This diagram shows the operational steps of a \texttt{CA}. In the first step, the \texttt{CA} must receive, study, and accept the terms and conditions, and then lock a certain amount of money as collateral in the IdentityChain smart contract. In the second step, a vote is conducted on the membership of this \texttt{CA}. If accepted, the \texttt{CA} can authenticate users. In the third step, the \texttt{CA} must notify 6 months in advance to exit the system and transfer its database. Then, the locked money is released. During these steps, the \texttt{CA} might be expelled from the system, but this is not shown because it is not part of the regular process.}
    \label{CA_steps}
\end{figure}

The tasks of a CA include, but are not limited to, conducting KYC procedures and issuing certificates for users, as well as securely storing encrypted user information. \texttt{CA}s are rewarded for each certificate they issue. However, \texttt{CA}s can also face penalties if a reveal proposal is accepted and it is found that the \texttt{CA}, who provided the certificate for the user in the proposal, has either lost the encrypted data or failed to properly perform their KYC duties. Each \texttt{CA} is assigned a credit score, which determines their ability to issue certificates. This score is based on the collateral provided by the \texttt{CA}, as well as their track record of adding users to the system and maintaining honesty over time. The schematic of the three aforementioned steps is shown in Figure \ref{CA_steps}.

\texttt{User}: In order to utilize IdentityChain, each user is required to undergo the KYC process with one of the system's \texttt{CA}s. Upon completion of the KYC process, users gain access to the benefits of IdentityChain's single-sign-on system. These advantages encompass a broad spectrum of \texttt{Website}s that leverage IdentityChain as their KYC third party, allowing \texttt{User}s to undergo KYC just once within the IdentityChain ecosystem. 

An \texttt{User} can join the system through several steps:
\begin{itemize}
\item First step: The \texttt{User} visits a \texttt{CA} to undergo the necessary KYC procedures. This KYC process adheres to established international standards. Upon successful completion of the KYC process, the \texttt{CA} issues a certificate for the \texttt{User} by which the \texttt{User} can register the generated public key as an anonymous and accountable account in the blockchain. 
\item Second step: The \texttt{User} can create a new blockchain account and fund it with IdentityChain tokens to cover the \texttt{CA} costs, burning some tokens to register a new anonymous account. They can then generate the necessary materials to register this account as a registered anonymous account in the IdentityChain smart contract using the certificate obtained in the previous step. 
\item Third step: The \texttt{User} employs our system's single-sign-on feature. Platforms integrating IdentityChain's single-sign-on can verify a \texttt{User}'s KYC status by querying IdentityChain. This verification process involves simply invoking a read function within the smart contract. Every six months, the \texttt{User} is required to consult either the current or a new \texttt{CA} to obtain a renewed certificate. 
\end{itemize}

\begin{figure}[H]    
    \centering
    \includegraphics[width=1\textwidth]{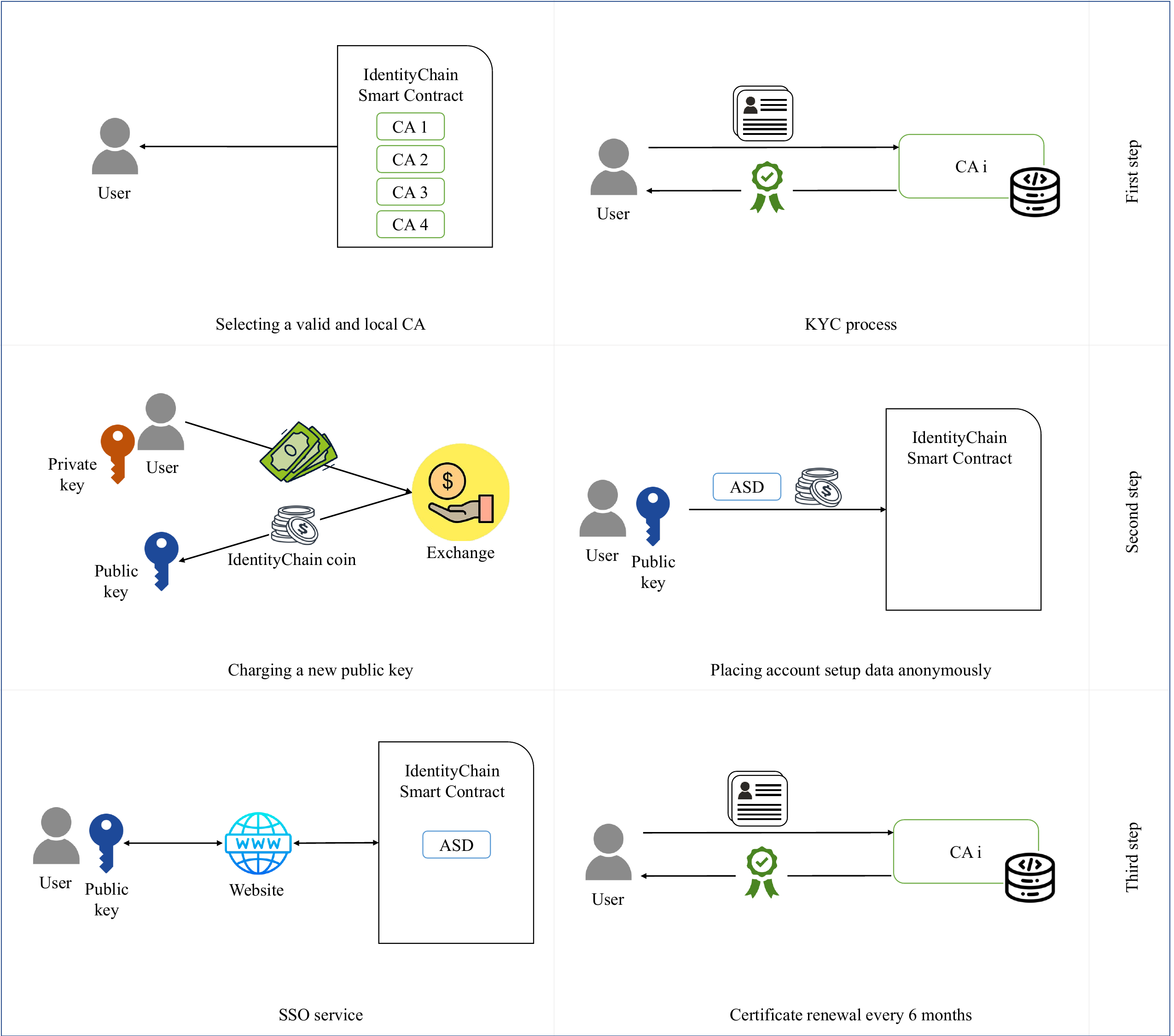} 
    \caption{
In this figure, the steps that the \texttt{User} takes are shown. In the first step, the \texttt{User} selects their geografical area's relevant \texttt{CA} from among the verified \texttt{CA}s. Then, they go through the KYC process to obtain a certificate. In the second step, a new public key and private key is generated. It is charged with the IdentityChain token and then sent anonymously, along with a fee, to the smart contract. This fee is given to the corresponding \texttt{CA}. In third step, the \texttt{User} visits a \texttt{Website}. The \texttt{Website} allows the \texttt{User} access by checking the smart contract. The KYC process must be repeated every six months. It is possible for the \texttt{User}'s identity to be disclosed by the decision of \texttt{SC}, but it is not shown in the figure as it is not part of the normal process.}
    \label{User_steps}
\end{figure}

If a \texttt{User} does not engage in any suspicious activities within the system, they can freely utilize it without interruption. However, in the event of suspicious activities, platforms that utilize IdentityChain as their KYC service have the option to submit a Revealing Proposal (\texttt{RP}) to ascertain and verify the real identity of a \texttt{User}. At this juncture, Supreme Committee (\texttt{SC}) members will vote on the \texttt{RP}. If the \texttt{RP} is accepted, the \texttt{User}'s identity will be accessible to the platform that submitted the \texttt{RP}. This mechanism ensures that the privacy and security of \texttt{User}s are upheld while allowing for necessary scrutiny in cases of suspicion. The schematic of the three aforementioned steps is shown in Figure \ref{User_steps}.

% Background
%%%%%%%%%%%%%%%%%%%%%%%%%%%%%%%%%%%%%%%%%%%%%%
% \section{Theoretical Principles} 
\section{Privacy and accountability} 
\label{Privacy_and_accountability}
\subsection{Prerequisites}
In this part we provide informal definitions of the building blocks used in the following protocols. The precise definitions are elaborated on in \cite{damgaard2021balancing}. 

\begin{definition}
Pseudo-random Functions ($\mathrm{PRF}$). A pseudo-random function $\mathrm{PRF}_K(x)$ is a deterministic function that maps the input $x$ to seemingly random output values, thereby exhibiting properties of randomness. However, unlike truly random functions, PRFs are computationally efficient and produce outputs that are indistinguishable from truly random values to any adversary without knowledge of the secret key ($K$) used to generate them. Various methods have been proposed for generating PRFs \cite{katz2007introduction}.
\end{definition}

\begin{definition}
Blind Signature Scheme. Blind signature schemes are cryptographic protocols that allow a user to obtain a valid signature on a message without revealing the content of the message to the signer. In a blind signature scheme, the user blinds the message before sending it to the signer for signature. The signer then signs the blinded message, producing a signature that is valid for the original, unblinded message. After receiving the signature, the user can unblind it to obtain a valid signature on the original message. In other words, if a person enters this algorithm as a signer, they can sign any message without seeing it \cite{okamoto2006efficient, schroder2012security}. 
\end{definition} 

\begin{definition}
Zero-knowledge proofs (ZKPs) are a cryptographic technique that allows someone (the prover) to convince someone else (the verifier) that a statement is true, without revealing anything about the statement itself. It's like proving the balance of an account is enough to make a transaction without revealing the balance. In the ZKPs, the prover sends a string, called proof, to the verifier to convince her about the claim \cite{bootle2016efficient}.    
\end{definition}

\begin{definition}
We can encrypt a message using someone's public key. In this case, only that person, who holds the corresponding private key, can decrypt the message. Now, suppose we have \( n \) individuals, each with a public key and a private key, and we want to encrypt a message \( m \) in such a way that any subset of \( d+1 \) individuals can decrypt the message and any subset with fewer than \( d+1 \) individuals cannot obtain any information about \( m \). This process is known as threshold encryption. To encrypt the message, it suffices to use the public keys of the \( n \) individuals. Then, for decrypting the encrypted message, each of the \( d+1 \) individuals performs their share of the decryption on the encrypted string. It is enough for one of them or someone else to combine these shares to retrieve the original message \( m \). Note that the decryption does not need to be performed in sequence; everyone can perform their share of the decryption in parallel \cite{boneh2018threshold}. 
\end{definition}

\subsection{Components and data objects} 
In this section, we explain the role of each component in the IdentityChain system. The roles are interconnected and may seem unclear at first, but everything will become clear when we review the protocols at the end.

\begin{itemize}
\item \texttt{User}s: Each \texttt{User} on the blockchain aims to open one or more verified and anonymous accounts and perform transactions. We assume these \texttt{User}s have a way to conduct legal identification in their country, such as with a passport or other personal documents (\texttt{User docs}). Thus, \texttt{User}s must first register with a Certification Authority (\texttt{CA}). This \texttt{CA} acts as a trusted third-party that verifies a \texttt{User}'s identity and grants them a certificate denoted by \texttt{User cert}. \texttt{User}s remain anonymous as long as they are honest and do nothing wrong. 

\item Certificate Authority (\texttt{CA}): 
The \texttt{User}s on the Blockchain rely on a trusted Certificate Authority (\texttt{CA}) to verify their real-world identity and issue a certificate. This certificate allows users to open accounts and conduct transactions anonymously. While the blockchain records which \texttt{CA} authorized an account, the true identity of the user remains hidden, even from the \texttt{CA} itself. This creates a system where everyone can verify the legitimacy of an account based on the trusted \texttt{CA}, yet user privacy is maintained. \texttt{CA} receives the personal documents of \texttt{User}s, saves a record of them, extracts the attributes of \texttt{User} and forms the attribute list \texttt{AL}. Finally, it signs a certificate for the \texttt{User}. We can have multiple \texttt{CA}s in the system. For a new \texttt{CA} to join, it should lock a deposit and get a confirmation from the Supreme Committee denoted by \texttt{SC}. For a \texttt{CA} to free its deposit and leave the system, it should transfer all of its records to a new verified \texttt{CA}, then get a confirmation from the Supreme Committee \texttt{SC}.

\item Supreme Committee (\texttt{SC}): Imagine a system where \texttt{User}s can interact anonymously on a blockchain, like a pseudonym on a forum. Members of \texttt{SC} are like trusted authorities who can reveal each \texttt{User}'s real identity under certain conditions. We can think of it like a safety deposit box with multiple keys. Each member of \texttt{SC} has one key. To open the box and reveal the \texttt{User}'s identity, at least \( d+1 \) of those keys need to be used together. This threshold ensures that a \texttt{User}'s identity is not accidentally or easily revealed. Law enforcement or other authorized entities might be some of these key holders (\texttt{SC} members). They can only gain access to the identity and the personal documents of a \texttt{User} if at least \( d+1 \) of them agree to do so. In addition to handling complaints about users and blocking certain accounts, the \texttt{SC} is also responsible for registering and evaluating \texttt{CA}s. If necessary, they can burn part of the \texttt{CA}s' collateral. Furthermore, they have the authority to add or expel SC members; in the case of expulsion, the expelled members' stakes will be burned. All users whose \texttt{CA} has been removed from the system, or who have a specified number of associated \texttt{SC} members no longer active in the system, must reapply for certification. 

\item \texttt{Website}: \texttt{Website}s utilize IdentityChain to easily register their users. In other words, when they see an account public key on the \texttt{User}s board, they can be assured that the user is registered. \texttt{Website}s do not incur costs with IdentityChain, but in order to file a complaint about an account's status with one or more \texttt{SC} members, they must hold a minimum amount of tokens from IdentityChain in their account.

\item \texttt{User docs}: Personal documents of a \texttt{User}. These may include a photocopy of the passport, a photocopy of the birth certificate, address, phone number, etc.

\item \texttt{IDcredPUB}: This is a public key stored in the \texttt{CA}'s records and is not disclosed to the public. If the user misbehaves in the system or engages in illegal activities, this public key will be revealed. Subsequently, the \texttt{CA} can access the \texttt{User docs} of the respective user.

\item \texttt{IDcredSEC}: This is the secret key associated with \texttt{IDcredPUB}. The \texttt{User} uses this key in the blind signature process that produces \texttt{ERegID}.

\item $K$: This is a key that specifies the $\textrm{PRF}$ used to generate unique IDs for \texttt{User} accounts in the blockchain, denoted by $\texttt{RegID}_\text{ACC}$. $K$ is chosen by the \texttt{User} at registration time. 

\item $\textrm{PRF}$: This function has two inputs: \( K \) and \( x \). \( K \) is an input that is fixed for each user, and \( x \) is an input that must be one of the numbers from $1$ to $\text{MaxACC}$. This function produces a random output that is used for unique user IDs on the blockchain denoted by $\texttt{RegID}_\text{ACC}$.

\item \texttt{AL}: This is a list of attributes that the user possesses and which the \texttt{CA} has checked and verified based on personal documents \texttt{User docs}. This list is confidential with the \texttt{CA}, and the \texttt{User} does not disclose it to the public. Instead, the \texttt{User} proves the possession of some of these attributes in zero-knowledge when necessary.

\item $\texttt{RegID}_\text{ACC}$: This is an account registration ID. This is defined to be $\texttt{RegID}_\text{ACC} = \text{PRF}_K(x)$ where $K$ is a key held by the \texttt{User} and signed by the \texttt{CA}, and where the account in question is the $x$’th account opened by the \texttt{User} based on a given $\texttt{User cert}$. If $\texttt{User}$ behaves honestly, then \texttt{RegID}\text{ACC} is unique for the account, and $x \leq \text{MaxACC}$. The latter condition is enforced by the proof $\pi$, the former can be checked publicly. 

\item $\text{pk}_{\texttt{CA}}$: The public key of the Certificate Authority (\texttt{CA}). Within the \texttt{ASD} stored on the blockchain, $\text{pk}_{\texttt{CA}}$ is visible to everyone, allowing them to identify the \texttt{CA} associated with that user. 

\item $\text{pk}_{\text{ACC}}$: This is the unique public key of an account used to generate transactions on the blockchain. It is created by a registered user and included in the \texttt{ASD}, then stored on the blockchain to establish a new account. \emph{No one} can associate this public key with the user's other accounts or documents as long as the user remains active and not revoked. 

\item $[\text{pk}_i]_{i \in \texttt{SC}}$: The public keys of the Supreme Committee (\texttt{SC}) members. These public keys are utilized in two scenarios: when the user needs to generate \texttt{EID}, which encrypts \texttt{IDcredPUB}, and \texttt{ERegID}, which encrypts $K$. 

\item $\sigma$: This is the signature on (\texttt{IDcredSEC}, $K$, \texttt{AL}) generated during the User Registration Protocol using a blind signature. It can be verified using $\text{pk}_{\texttt{CA}}$. 

\item \texttt{EID}: This is a threshold encryption $\texttt{EID} = \text{TEnc}_{n,d} ([\text{pk}_i]_{i \in {\texttt{SC}}}, \texttt{IDcredPUB})$, where any subset of size $d + 1$ of \texttt{SC} are able to decrypt $\texttt{EID}$ and obtain $\texttt{IDcredPUB}$. This is used for anonymity revocation. 

\item $\text{MAX}_{\text{ACC}}$: This is the maximum number of accounts that each registered user is permitted to create on the blockchain. 

\item $\pi$: This is a zero knowledge proof that can be checked using $\text{pk}_{\texttt{CA}}$ and verifies that \texttt{ASD} can only be created by a \texttt{User} that has obtained a \texttt{User cert} from \texttt{CA}, such that $P(\texttt{AL}) = \text{True}$, where \texttt{User} knows the secret keys corresponding to $\text{pk}_{\text{ACC}}$, as well as $\texttt{IDcredSEC}$ corresponding to the $\texttt{IDcredPUB}$ that was presented to the \texttt{CA}, and where $\texttt{RegID}_\text{ACC}$, $\texttt{EID}$ and $\texttt{ERegID} = \text{TEnc}_{n,d} ([\text{pk}_i]_{i \in {\texttt{SC}}}, K)$ are correctly generated. 

\item $P$: This is a policy that specifies certain information about the attribute list \texttt{AL}. 

\item $\texttt{ERegID}$: This is the encrypted form of $K$, represented as $\texttt{ERegID} = \text{TEnc}_{n,d}([\text{pk}i]{i \in \texttt{SC}}, K)$, and it is generated by the \texttt{User}.

\item \texttt{User cert}: After completing registration with a \texttt{CA}, the user receives a certificate called \texttt{User cert}. This certificate is a tuple consisting of (\texttt{IDcredPUB}, \texttt{IDcredSEC}, $K$, \texttt{AL}, $\sigma$). It is important to note that the user does not publish all parts of the \texttt{User cert}. 

\item Account Setup Data (\texttt{ASD}): Given a \texttt{User cert}, a user can create new accounts and post the corresponding \texttt{ASD} on the ledger. This \texttt{ASD} is a tuple consisting of $(\texttt{RegID}{\text{ACC}}, \texttt{EID}, [\text{pk}_i]_{i \in \texttt{SC}}, \text{pk}_{\texttt{CA}}, \text{pk}_{\text{ACC}}, P, \pi)$. The public key $\text{pk}{\texttt{CA}}$ belongs to the identity provider who signed the \texttt{User cert} used for this account. 

\item \texttt{CA}'s record (\texttt{CA record}). This is the data record that the \texttt{CA} stores after a \texttt{User} has registered. It is the tuple $(\texttt{User docs}, \texttt{IDcredPUB}, \texttt{AL}, \texttt{ERegID}, [\text{pk}_i]_{i \in {\texttt{SC}}})$.

\item Revealing Proposal (\texttt{RP}): Each \texttt{SC} member oversees the network. Additionally, they may receive complaints or requests from \texttt{Website}s or external entities to disclose the identities of the owner of certain accounts. In such cases, each \texttt{SC} member can submit a proposal to reveal the identity of a \texttt{User} who owns an account in Revealing Proposal board. This proposal is called a Revealing Proposal (\texttt{RP}) and includes the \texttt{ASD} associated with the \texttt{User}'s account. Other \texttt{SC} members vote on it, and if approved, the protocol Revoke Anonymity of Account is executed. 

\item The relation $R$: This relation accepts a set of public and private inputs. As illustrated in Figure \ref{fig:circuit}, it conducts computations and produces either True or False. Since the \texttt{User} is unable to transmit private inputs to the blockchain, they calculate the zero-knowledge proof $\pi$ and transmit it alongside the public inputs during the Protocol Create New Account.
\end{itemize}

\subsection{Protocols}
The following are the main protocols in our design. 

\begin{itemize}
\item \texttt{User} Registration \\
This protocol occurs between a \texttt{CA} and a \texttt{User} who possesses a key pair $(\texttt{IDcredSEC}, \texttt{IDcredPUB})$. At the conclusion of the protocol, the \texttt{User} receives a \texttt{User cert} and the attribute list \texttt{AL}, and the \texttt{CA} obtains a \texttt{CA record} as described above. The \texttt{User} submits their personal documents to the \texttt{CA} and verifies their identity to the \texttt{CA} through non-cryptographic means. Specifically, the \texttt{CA} must confirm that the entity it is communicating with indeed has the personal documents $\texttt{User docs}$ and forms a list of attributes of \texttt{User} in a list denoted by $\texttt{AL}$. Additionally, the \texttt{CA} should verify that the attributes in $\texttt{AL}$ are accurate with respect to the \texttt{User}. 
The \texttt{User} also sends their public key $\texttt{IDcredPUB}$ and an encryption $\texttt{ERegID} = \text{TEnc}_{n,d} ([\text{pk}_i]_{i \in \texttt{SC}}, K)$, where $K$ is a $\textrm{PRF}$ key, to the \texttt{CA}. Subsequently, the \texttt{User} and \texttt{CA} engage in a blind signature scheme, enabling the \texttt{User} to receive the signature $\sigma$ on $(\texttt{IDcredSEC}, K, \texttt{AL})$ generated under the secret key $\text{sk}_{\texttt{CA}}$ of the \texttt{CA}. 
% Moreover, the \texttt{User} proves (cryptographically, in zero-knowledge) that they know $\texttt{IDcredSEC}$ corresponding to $\texttt{IDcredPUB}$, that the same $\texttt{IDcredSEC}$ was used in the blind signature, and that the encryption contains the same $K$ used in the blind signature scheme for the generation of $\texttt{ERegID}$. 
The \texttt{CA} stores $\texttt{CA record} = (\texttt{User docs}, \texttt{IDcredPUB}, \texttt{AL}, \texttt{ERegID}, [\text{pk}_i]_{i \in \texttt{SC}})$.

\item Create New Account \\
A \texttt{User} wishes to create an account that complies with a policy $P$ (e.g., being above 18, residing in country X, etc.). They use a \texttt{User cert}, a policy $P$, and the set of public keys $[\text{pk}_i]_{i \in \texttt{SC}}$ of the $\texttt{SC}$ members. At the end, \texttt{User} produces some \texttt{ASD} that can be posted to the blockchain. They also need to store a secret key $\text{sk}_{\text{ACC}}$ specific to the account. The protocol proceeds as follows: \texttt{User} generates an account key pair $(\text{pk}_{\text{ACC}}, \text{sk}_{\text{ACC}})$ and an encryption of their public identity credential $\texttt{IDcredPUB}$ under the public key of the Supreme Committee members $[\text{pk}_i]_{i \in {\texttt{SC}}}$, i.e., $\texttt{EID} = \text{TEnc}_{n,d} ([\text{pk}_i]_{i \in {\texttt{SC}}}, \texttt{IDcredPUB})$. Next, \texttt{User} calculates $\texttt{RegID}_{\text{ACC}} = \textrm{PRF}_K(x)$, assuming this is the $x$'th account opened using the \texttt{User cert} provided. At last, \texttt{User} produces a zero-knowledge proof $\pi$ for the relation $R$ that outputs $\text{True}$ if: 
\begin{enumerate}
    \item $\sigma$ is a valid signature under $\text{pk}_{\texttt{CA}}$ for a message of the form $(\texttt{IDcredSEC}, K, \texttt{AL})$.
    \item $\texttt{AL}$ satisfies the policy, i.e., $P(\texttt{AL}) = \texttt{True}$.
    \item $\texttt{RegID}_{\text{ACC}} = \textrm{PRF}_K(x)$ for some $x \leq \text{Max}_{\text{ACC}}$.
    \item $\texttt{EID} = \text{TEnc}_{n,d}([\text{pk}_i]_{i \in \texttt{SC}}, \texttt{IDcredPUB})$.
    \item $(\text{pk}_{\text{ACC}}, \text{sk}_{\text{ACC}})$ is a valid key pair.
\end{enumerate}

The arithmetic circuit for the aforementioned relation is illustrated in Figure \ref{fig:circuit}. It should be noted that $\texttt{ASD} = (\texttt{RegID}_{\text{ACC}}, \texttt{EID}, [\text{pk}_i]_{i \in \texttt{SC}}, \text{pk}_{\texttt{CA}}, \text{pk}_{\text{ACC}}, P, \pi)$ represents the tuple of public inputs, the policy $P$, and the proof $\pi$ for the relation $R$.

\begin{figure}[H]
    \centering
    \includegraphics[width=1\textwidth]{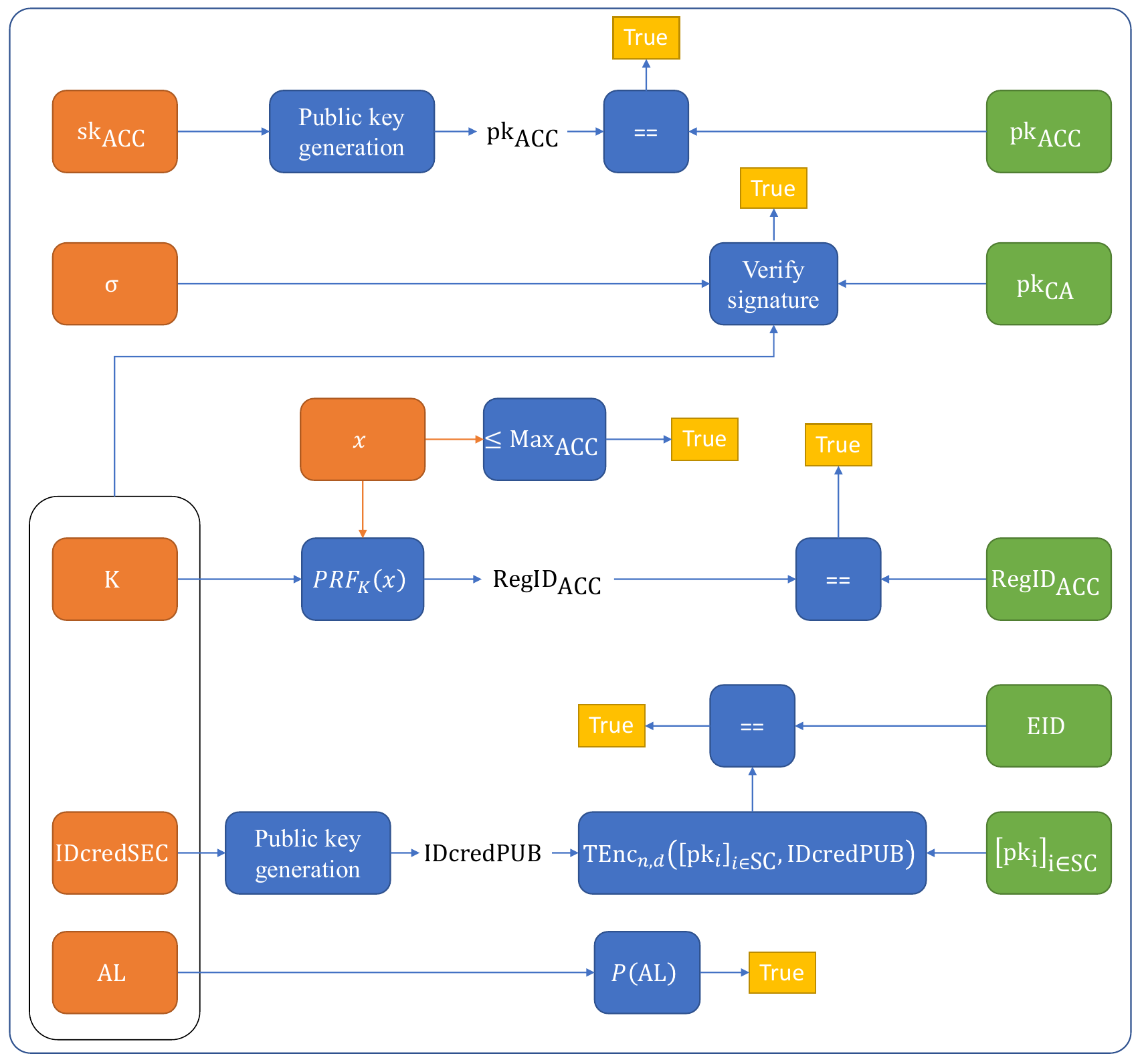} % Replace 'example-image' with the filename of your image
    \caption{The arithmetic circuit of the relation $R$ which is to be verified in the Protocol Create New Account. The prover (\texttt{User}) demonstrates to the public that all six outputs of the arithmetic circuit evaluate to $\text{True}$. In this illustration, public inputs are denoted by green boxes, private inputs are represented by orange boxes, and functions are depicted in blue boxes (requires more elaborations).} 
    \label{fig:circuit}
\end{figure}

\item 
Revoke Anonymity of Account \\ 
Revoking the anonymity of an account requires the collaboration of at least $d+1$ out of the $n$ Supreme Committee (\texttt{SC}) members whose public keys are stored in the Account State Data (\texttt{ASD}) of the respective account, posted on the blockchain during its creation. They work in conjunction with the Certificate Authority (\texttt{CA}) with whom the user registered. The input for this process is an account identifier $\texttt{RegID}_{\text{ACC}}$, and the outputs are the public keys $\text{pk}_\text{ACC}$ of all the \texttt{User}'s accounts on the blockchain, along with the \texttt{User docs} containing real identity and additional information about the \texttt{User}.

The protocol unfolds as follows: When an account $\texttt{RegID}_{\text{ACC}}$ requires its anonymity to be revoked, the \texttt{SC} members locate the corresponding \texttt{ASD} on the blockchain, collaborate to decrypt \texttt{EID}, and acquire \texttt{IDcredPUB}. The registration information within \texttt{ASD} also includes the public key $\text{pk}_\texttt{CA}$ used to register \texttt{IDcredPUB}. These \texttt{SC} members then contact the $\texttt{CA}$ to locate the $\texttt{CA record} = (\texttt{User docs}, \texttt{IDcredPUB}, \texttt{AL}, \texttt{ERegID}, [\text{pk}_i]_{i \in \texttt{SC}})$ associated with the decrypted \texttt{IDcredPUB}. This record contains the $\texttt{User docs}$, allowing the \texttt{SC} members to identify the \texttt{User}. Additionally, the record includes \texttt{ERegID}.

These $d+1$ \texttt{SC} members collaborate again to derive the $\textrm{PRF}$ key $K$ and subsequently generate all values $\textrm{PRF}_K(x)$ for $x = 1, \ldots, \text{MaxACC}$, representing the possible unique IDs of accounts that the $\texttt{User}$ could have created. Consequently, the \texttt{SC} members can identify all of the \texttt{User}'s accounts. Finally, the \texttt{SC} members collectively determine the next steps based on this information.

\end{itemize}

%%%%%%%%%%%%%%%%%%%%%%%%%%%%%%%%%%%%%%%%%%%%%%
\section{System Setup}
\label{System_Setup}
In the previous section, we reviewed the components and protocols through which a user could register in the system, operate anonymously, and, if they committed an offense, have their identity and all their accounts revealed. For the system to function effectively, other details are also necessary. For example, we need to understand how Certificate Authorities (CAs) join the system, how they are compensated, how they exit the system, and what violations they might commit. Similar questions arise regarding the members of \texttt{SC} and the websites. In this section, we aim to address these questions and provide a comprehensive overview of the functioning of these components. To better explain how the various components interact with one another, we utilize a framework of boards for information sharing. Each board contains specific information, with certain individuals having write access and others having read access. For these boards to be effectively implemented and function smoothly, several smart contracts need to be developed in the background. For the sake of simplicity, we will not delve into the details of these smart contracts. In the following, we introduce the boards we use and explain what information is contained within each one, who can edit each board, and who can view the information on them. 

\begin{itemize}
\item \texttt{SC} members board \\
Explanation: This board shows the registered \texttt{SC} members. If an \texttt{SC} member behaves improperly, for example by abstaining from voting or not participating in decryption, the other members can decide to burn their stake and expel them. \\
Write access: The majority of the already registered \texttt{SC} members.  \\
Read access: Everyone on the blockchain. \\
Board items: 
	\begin{itemize} 
	\item 
	Public keys of registered \texttt{SC} members.
	\item 
	Stake of each \texttt{SC} member.
	\item 
	Burned stake.
	\item 
	Expelled members. 
	\item 
	Record of write/edit transactions.
	\end{itemize} 
Figure \ref{SC_members_board} depicts an illustration of this board.	
	
\begin{figure}[H]    
    \centering
    \includegraphics[width=1\textwidth]{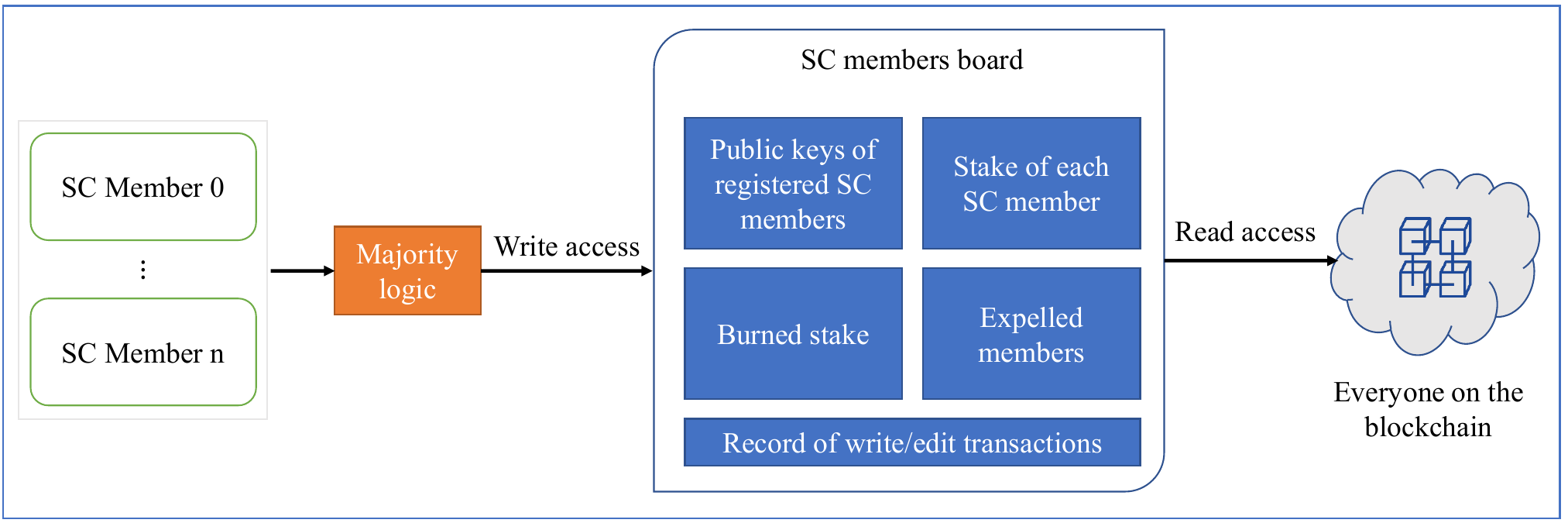} 
    \caption{The SC members board that can be implemented using smart contracts in IdentityChain.}
    \label{SC_members_board}
\end{figure}   

\item Revealing Proposal board \\
Explanation: On this board, each member of the \texttt{SC} can propose revealing the identity of the owner of a misbehaving account by presenting the \texttt{ASD} of the account on the blockchain. Each \texttt{ASD} contains the public key of $n$ \texttt{SC} members. All $n$ members are required to participate in the vote to decide whether they agree with the proposal. If at least $d+1$ members agree, they execute the protocol Revoke Anonymity of Account to uncover the \texttt{User docs} and post it on this board. \\
Write access: Each \texttt{SC} member. 
\\
Read access: All \texttt{SC} members. \\
Board items: 
	\begin{itemize} 
	\item 
	Awaiting proposals for voting (\texttt{ASD}, list of signed votes)
	\item 
	Running instances of the protocol Revoke Anonymity of Account (decryption shares, exchanged messages with \texttt{CA}s). 
	\item 
	A list of revealed \texttt{User docs} and the corresponding \texttt{User}'s accounts. 
	\item 
	Rejected proposals.
	\item 
	Record of write/edit transactions. 
	\end{itemize} 
Figure \ref{Revealing_Proposal_board} illustrates a schematic of this board. 
	
\begin{figure}[H]    
    \centering
    \includegraphics[width=1\textwidth]{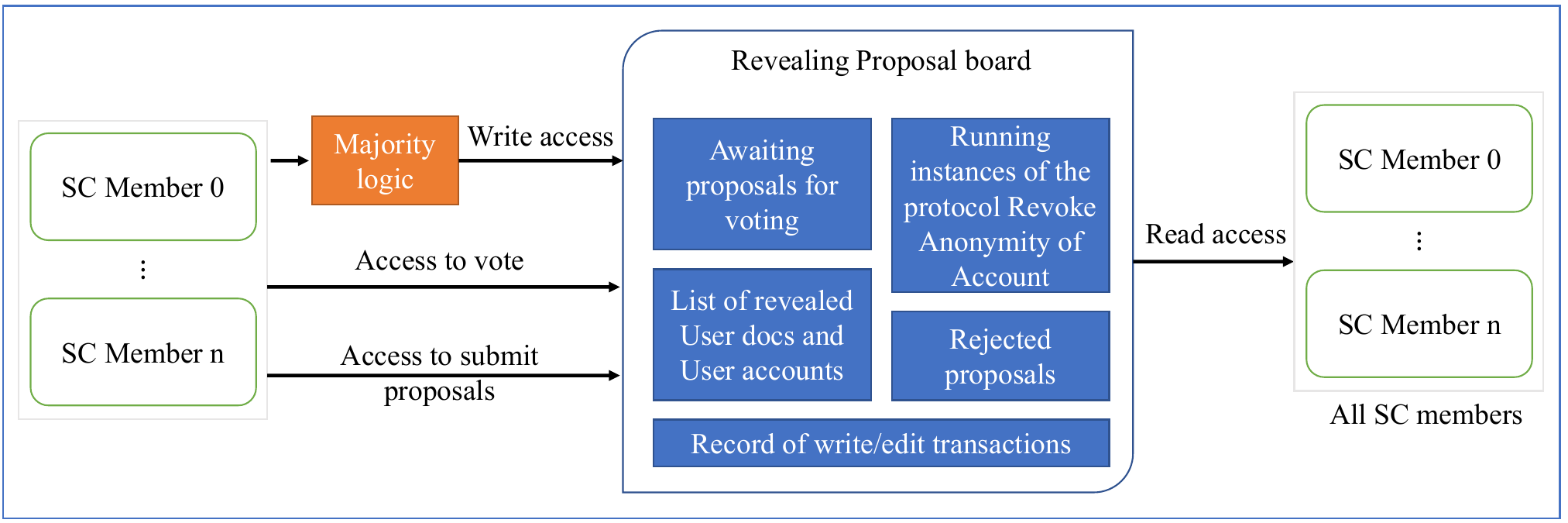} 
    \caption{In this figure, the proposals board is depicted. It's important to note that this board is not visible to all blockchain members. Consensus among SC members is achieved using this board.}
	\label{Revealing_Proposal_board}
\end{figure}   	
	
\item \texttt{CA}s board \\
Explanation: This board displays the public keys of \texttt{CA}s along with their performance scores and indicates how long each \texttt{CA} will remain active in IdentityChain. If a \texttt{CA} becomes inactive, the ASDs of all users who have its signature will become invalid, requiring these users to obtain new certificates from a new \texttt{CA}. \\
Write access: The majority of the already registered \texttt{SC} members.  \\ 
Read access: Everyone on the blockchain. \\
Board items: 
	\begin{itemize} 
	\item 
	List of registered \texttt{CA}s (their public keys and access link).
	\item 
	Score of each \texttt{CA}.
	\item 
	Exit time of each \texttt{CA}.
	\item 
	The operational scope of each \texttt{CA}: Each \texttt{CA }operates within a specific domain, such as a geographic region, and adheres to local regulations while also complying with the general system rules.
	\item 
	Record of write/edit transactions. 
	\end{itemize}  
Figure \ref{CAs_board} depicts an illustration of this board.	
	
\begin{figure}[H]    
    \centering
    \includegraphics[width=1\textwidth]{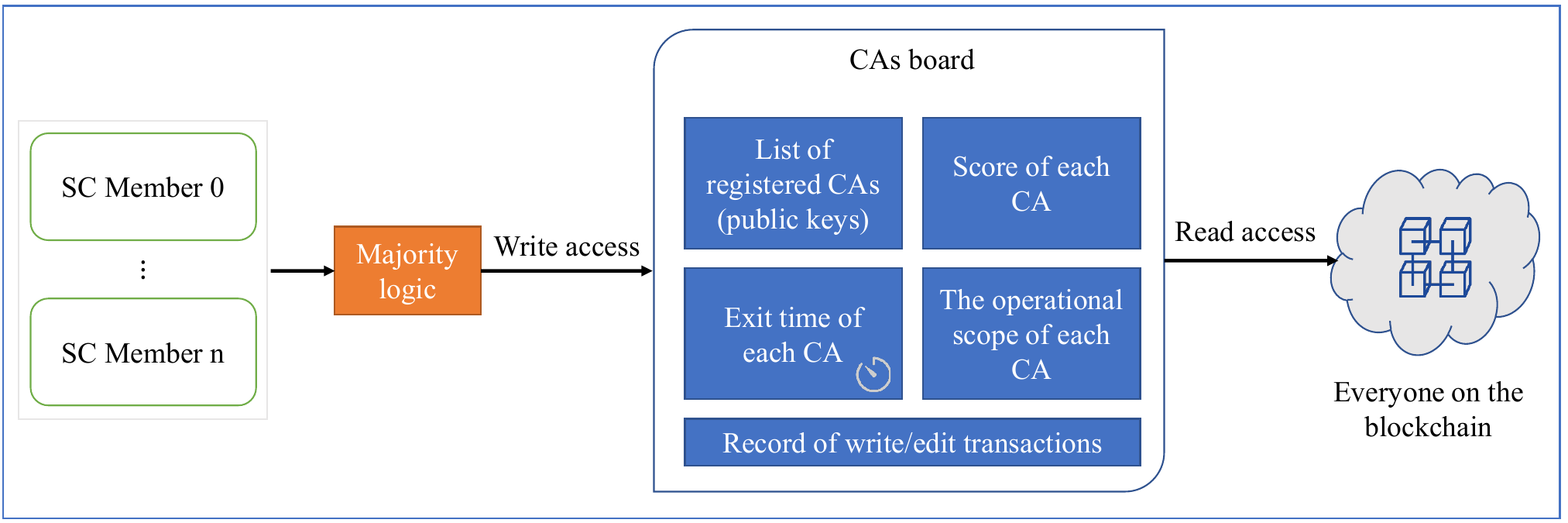} 
    \caption{The CA board is shown in this figure. This board is implemented in IdentityChain.}
    \label{CAs_board}
\end{figure}   		

\item \texttt{User}s board \\
Explanation: This board contains a list of \texttt{ASD}s from registered \texttt{User}s. \texttt{User}s on this list are allowed to add, or deactivate their own \texttt{ASD}s but do not have permission to edit others' information. To add a new \texttt{ASD}, the \texttt{User} must transfer a predetermined amount of tokens to the \texttt{CA} mentioned in the \texttt{ASD} and burn another predetermined amount of tokens from their account. Any information on this board can be edited with a transaction initiated by a majority of \texttt{SC} members. \\     
Write access: \texttt{Users}s (limited access explained above) and the majority of the already registered \texttt{SC} members.  \\ 
Read access: Everyone on the blockchain. \\
Board items: 
	\begin{itemize} 
	\item 
	List of \texttt{ASD}s. 
	\item 
	List blocked $\text{pk}_{\text{ACC}}$s. 
	\item
	List of deactivated accounts.  
	\item
	List of accounts requiring certificate renewal. This list includes accounts whose certificate renewal deadline has passed or whose CA has left the IdentityChain.
	\item
	Record of write/edit transactions. 
	\end{itemize} 
In Figure \ref{Users_board}, an illustration of this board is displayed.

\begin{figure}[H]    
    \centering
    \includegraphics[width=1\textwidth]{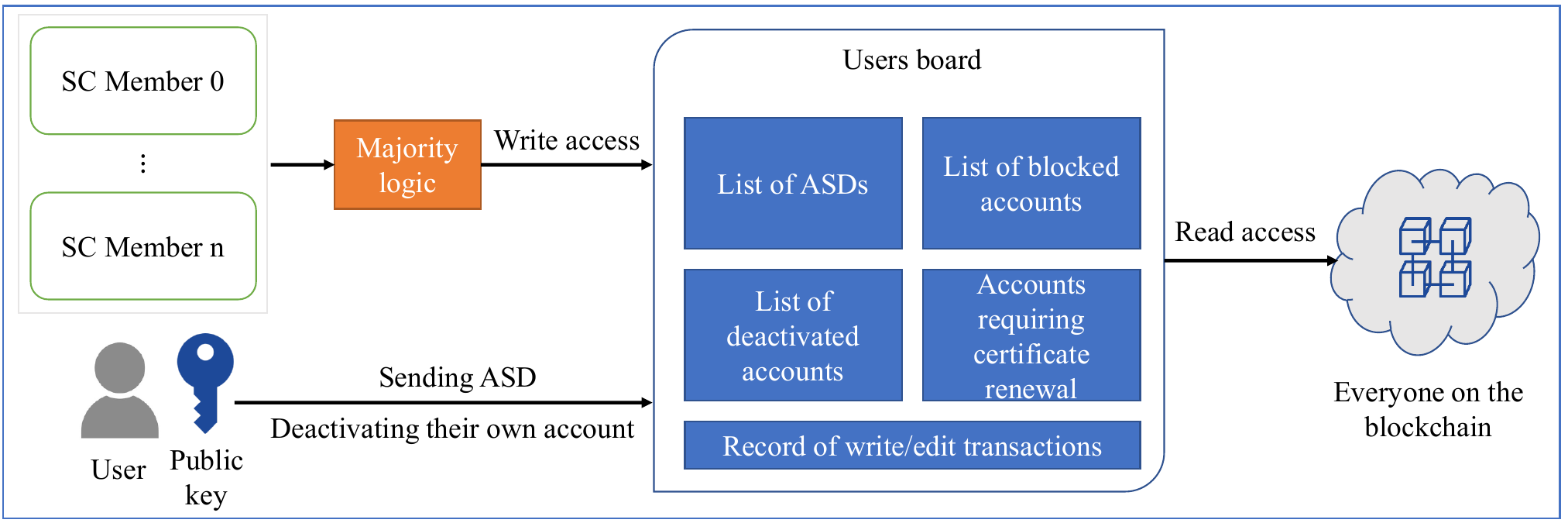} 
    \caption{The \texttt{User}s board is depicted in this figure. This board is the most active in the system, and \texttt{User}s need to submit \texttt{ASD} to this board to create an account.}
    \label{Users_board}
\end{figure}  	
\end{itemize}

\begin{figure}[H]
    \centering
    \includegraphics[width=1\textwidth]{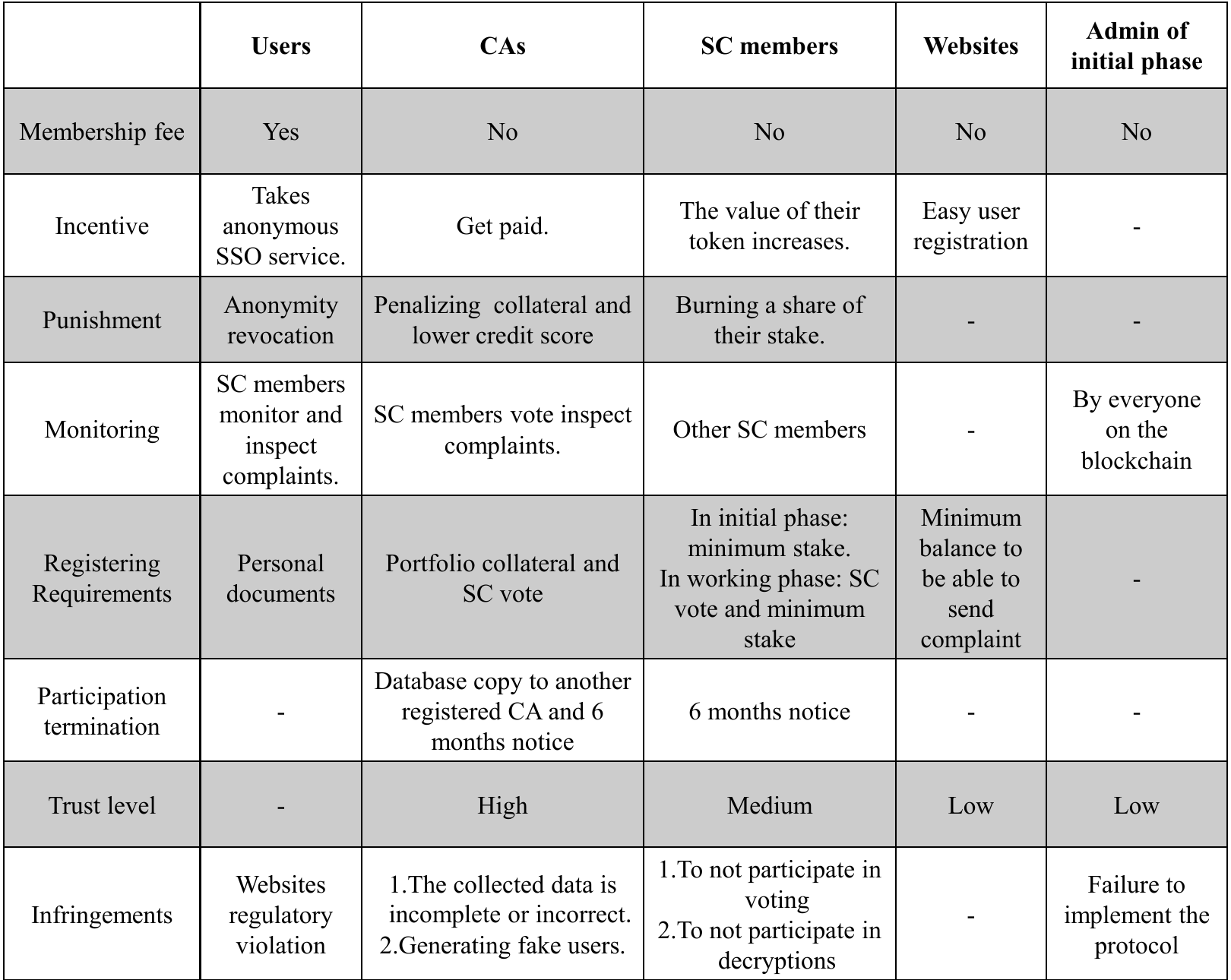} % Replace 'example-image' with the filename of your image
    \caption{Summary of IdentityChain setup information} 
    \label{fig:table}
\end{figure}

Summary of key information about the main components of the IdentityChain system is provided in Table \ref{fig:table}. This information includes whether they pay membership fees, their motivations in the system, costs incurred in case of errors, who monitors them, requirements for joining the system, termination of membership process, their trust level in the system, and types of errors they might commit.

\section{Implementation Vision} 
\label{Implementation}

\section{Conclusion}
\label{Conclusion}
In this article, we present the IdentityChain framework. This framework utilizes existing cryptographic solutions along with practical implementation scenarios in a blockchain environment to conduct the KYC (Know Your Customer) process for various websites in a confidential and secure manner. We have made efforts to clearly and precisely define the roles of all participants in this system. Additionally, there is flexibility to establish the necessary regulations for system governance in greater detail in the future without compromising the integrity of the overall system. The future directions of this white paper include its implementation and collecting feedback from users, regulatory bodies, and various individuals. Furthermore, there needs to be an examination of integrating IdentityChain with Layer 2 blockchain solutions, decentralized applications (DApps), data availability proofs, and more. 

\section*{Acknowledgement}
Thanks to Ali Rahimi for his assistance and support in writing this white paper. 

\bibliographystyle{ieeetr}
\bibliography{bibliography}
\end{document}